\documentclass[9pt,twocolumn,twoside]{pnas-arxiv}

\templatetype{pnasresearcharticle} 

\usepackage[utf8]{inputenc} 
\usepackage[T1]{fontenc}    
\usepackage{hyperref}       
\usepackage{url}            
\usepackage{booktabs}       
\usepackage{amsfonts}       
\usepackage{nicefrac}       
\usepackage{microtype}      
\usepackage{xcolor}         
\usepackage{outlines}
\usepackage{subcaption}
\usepackage{comment}

\usepackage[en-US,useregional,showdow]{datetime2}
\DTMlangsetup[en-US]{ord=raise}

\usepackage{amsthm}
\theoremstyle{definition}

\theoremstyle{plain}

\usepackage{graphicx}
\usepackage{longtable}
\usepackage{wrapfig}
\usepackage{rotating} 
\usepackage[normalem]{ulem}
\usepackage{amsmath}
\usepackage{amssymb}
\usepackage{capt-of}
\usepackage{hyperref}
\usepackage{amsthm}
\usepackage{xcolor}
\usepackage{soul}
\usepackage{physics}
\usepackage{slashed}
\usepackage{siunitx}
\usepackage{mathtools}
\usepackage{bm}
\usepackage{bbm}
\usepackage[capitalize]{cleveref}
\usepackage{multirow,diagbox}
\usepackage{nicematrix}
\usepackage{wrapfig}
\usepackage{array}
\usepackage{xr}
\usepackage{pdfpages}
\renewenvironment{boxed}
    {\tiny
    \begin{center}
    \begin{tabular}{|p{0.9\textwidth}|}
    \hline\\
    }
    { 
    \\\\\hline
    \end{tabular} 
    \end{center}
    }

\newcommand{\google}{Google, USA}
\newcommand{\cornell}{Department of Physics, Cornell University, USA}
\newcommand{\harvard}{Department of Physics, Harvard University, USA}
\newcommand{\stanford}{Department of Physics, Stanford University, USA}
\renewcommand{\mit}{Department of Physics, Massachusetts Institute of Technology, USA}
\newcommand{\jhu}{William H. Miller III Department of Physics and Astronomy, The Johns Hopkins University, Baltimore, MD, USA}
\newcommand{\bnl}{Condensed Matter Physics and Materials Science Division, Brookhaven National Laboratory, USA}
\newcommand{\flatiron}{Center for Computational Quantum Physics, Flatiron Institute, USA}
\newcommand{\cdf}{Collège de France, Paris, France}
\newcommand{\polytechnique}{CPHT, CNRS, Ecole Polytechnique, IP Paris, France}
\newcommand{\geneva}{DQMP, Université de Genève, Suisse}
\newcommand{\upsaclay}{Université Paris-Saclay, CNRS, CEA, Institut de physique théorique, France}
\newcommand{\cuny}{Physics Program and Initiative for the Theoretical Sciences, CUNY, USA}
\newcommand{\csi}{Department of Physics and Astronomy, College of Staten Island, CUNY, USA}
\newcommand{\harvardseas}{School of Engineering and Applied Sciences, Harvard University, USA}
\newcommand{\ewha}{Department of Physics, Ewha Womans University, Seoul, South Korea}

\begin{document}

\title{Expert Evaluation of LLM World Models: A High-T$_c$ Superconductivity Case Study
}

\author[a,1]{Haoyu Guo} 
\author[b,c,1]{Maria Tikhanovskaya}
\author[b]{Paul Raccuglia}  
\author[b]{Alexey Vlaskin} 
\author[b]{Chris Co} 
\author[b]{Daniel J. Liebling}  
\author[b]{Scott Ellsworth} 
\author[b]{Matthew Abraham}  
\author[b]{Elizabeth Dorfman} 

\author[d]{N. P. Armitage}  
\author[e]{Chunhan Feng}  
\author[e,f,g,h]{Antoine Georges}  
\author[e,i]{Olivier Gingras} 
\author[e]{Dominik Kiese}  
\author[j]{Steven A. Kivelson}  
\author[k,l]{Vadim Oganesyan}  
\author[a]{B. J. Ramshaw}  
\author[c]{Subir Sachdev}  
\author[m]{T. Senthil}  
\author[n]{J. M. Tranquada}  

\author[b,c,o]{Michael P. Brenner} 
\author[b]{Subhashini Venugopalan}  
\author[a,b,p,2]{Eun-Ah Kim}

\affil[a]{\cornell}
\affil[b]{\google}
\affil[c]{\harvard}
\affil[d]{\jhu}
\affil[e]{\flatiron}
\affil[f]{\cdf}
\affil[g]{\polytechnique}
\affil[h]{\geneva}
\affil[i]{\upsaclay}
\affil[j]{\stanford}
\affil[k]{\cuny}
\affil[l]{\csi}
\affil[m]{\mit}
\affil[n]{\bnl}
\affil[o]{\harvardseas}
\affil[p]{\ewha }

\leadauthor{Guo}

\significancestatement{Solving long-standing scientific problems requires researchers to navigate vast and complex research literatures with competing perspectives. While Large Language Models (LLMs) can aid in this challenging process in principle, reading scientific research literature to synthesize knowledge and identify conclusions supported by experimental evidence requires skills that go beyond everyday reading. Using high-temperature superconducitivity as an exemplar, we construct a unique dataset of questions and answers that probe expert knowledge of the literature. We then carry out systematic analysis of multiple LLM systems for answering questions at the level of an expert researcher. We find that curated data and multimodal retrieval significantly improve accuracy and depth. These findings inform the development of trustworthy AI tools to advance scientific progress.}
\equalauthors{\textsuperscript{1}H.G. contributed equally to this work with M.T.}
\correspondingauthor{\textsuperscript{2}To whom correspondence should be addressed. E-mail: eun-ah.kim@cornell.edu}


\begin{abstract}
 Large Language Models (LLMs) show great promise as a powerful tool for scientific literature exploration. However, their effectiveness in providing scientifically accurate and comprehensive answers to complex questions within specialized domains remains an active area of research. 
 Using the field of high-temperature cuprates as an exemplar, we evaluate the ability of LLM systems to understand the literature at the level of an expert. We construct
an expert-curated database of 1,726 scientific papers that covers the history of the field, and a set of 67 expert-formulated questions that probe deep understanding of the literature.  We then evaluate 
six different LLM-based systems for answering these questions, including both commercially available closed models and a custom retrieval-augmented generation (RAG) system capable of retrieving images alongside text. Experts then evaluate the answers of these systems against a rubric that assesses 
 balanced perspectives, factual comprehensiveness, succinctness, and evidentiary support.
Among the six systems two using RAG on curated literature outperformed existing closed models across key metrics, particularly in providing comprehensive and well-supported answers.  
 We discuss promising aspects of LLM performances as well as critical short-comings of all the models. The set of expert-formulated questions and the rubric will be valuable for assessing expert level performance of LLM based reasoning systems.
\end{abstract}

\dates{This manuscript was compiled on \today}

\maketitle

\thispagestyle{firststyle}
\ifthenelse{\boolean{shortarticle}}{\ifthenelse{\boolean{singlecolumn}}{\abscontentformatted}{\abscontent}}{}




\dropcap{T}here is a structural problem that impedes progress on long-standing scientific problems;  while the body of accumulated wisdom in the field contains valuable information, the sheer volume of literature makes exploiting this knowledge base extremely difficult for a new generation, even when they come armed with potentially game-changing insights, methodologies, or new information.  When a problem has remained interesting but only partially solved despite decades of work, only experts with long engagement with the field may appreciate what has been established, including reasons for abandoning seemingly promising lines of thought. At some point, it becomes impossible for a new generation to build on the body of literature from a fresh perspective, simply because it is difficult to acquire a comprehensive and critical understanding of what has come before. There is an opportunity here for LLMs to enable progress.

Ideally, a curious student would have an objective expert panel available on demand, answering researchers' questions in a trustworthy and comprehensive fashion. The goal of this paper is to evaluate whether Large Language Model  (LLM) guided assistants can accomplish this goal.  Within the experimental sciences, for a researcher to trust an answer, it must be 
grounded in experimental evidence, including data visualization of the relevant figures in the literature. When experimental results are challenging to reconcile, not because of reproducibility  but because existing theoretical frameworks place the results at odds with each other, the complexity in perspectives should be acknowledged. For expert researchers,  early experiments can have outsided importance, even if the experimental techniques are classic. Other early experiments or the conclusions drawn from them may have been later found to be misguided. Hence, the assistant should present the implications of the experiment in the context of its timing, specifically how it supports or counters previous observations and whether it is in harmony with other contemporary observations, as determined from a known theoretical framework.  
Perhaps the most valuable quality to seek in an ideal assistant is sound and critical judgment, enabling them to see beyond the author's bias and interpretation to extract objective facts. 


Here we evaluate the ability of LLM systems to serve as such an assistant in a particular problem of great scientific and technological importance: high-temperature superconductivity (HTS).   
The unexpected discovery in 1986 \cite{JGBednorz1986} of superconductivity at unprecedentedly high-temperatures in ceramic materials made of copper, oxygen, and various other elements had a singular and profound impact on condensed matter physics. 
Soon after this original discovery of high-temperature (high-$T_c$) superconductivity in what is now called the 214 (La$_{2}$CuO$_4$–based) family, additional structural families -- most notably Y-Ba-Cu-O (YBCO) and the Bi–Sr–Ca–Cu–O (BSCCO) series -- were also found to exhibit high-$T_c$ superconductivity \cite{MKWu1987a,HMaeda1988}. Many further families were identified in subsequent years, including Tl- and Hg-based cuprates as well as the electron-doped compounds \cite{ZZSheng1988,SNPutilin1993,takagi_superconductivity_1989}, thus establishing both the challenges and appeals of the field.
There is a diversity in the material landscape as each family of materials exhibits a rich set of phenomena upon changes in temperature, magnetic field, crystal structure, and charge carrier concentration. Moreover, these cuprate materials exhibit strange and unusual behavior even in the metallic state at temperatures above their superconducting transition temperatures. There are quantitative and qualitative variations in observed phenomena in this high-dimensional parameter space of materials explored by a large community using various samples over the span of four decades.
However, discerning what observations are specific to a particular sample, a particular subclass of materials, or a particular family, rather than being universal phenomena is challenging without a comprehensive understanding of the literature.

Over the decades, the scientific community has amassed a vast body of experimental data, dispersed across thousands of publications. Nevertheless, we still do not understand how to find a new high-$T_c$ superconductor or how to reconcile the many seemingly contradictory phenomena observed in this material class. 
The long-standing puzzles invited an ever-increasing list of experimental probes to tackle the problem from different angles, each revealing new facets of the problem: angle-resolved photoemission spectroscopy (ARPES), scanning tunneling microscopy (STM), neutron scattering, resistivity, nuclear magnetic resonance (NMR), nuclear quadrapolar resonance (NQR), THz spectroscopy, optical conductivity, thermal transport, specific heat, muon spin rotation ($\mu$SR), x-ray scattering, electron energy loss spectroscopy (EELS), raman scattering, magnetization, angle-dependent magnetoresistance (ADMR), and more.
A satisfactory theory must synthesize and reconcile experimental facts gathered from many complementary probes rather than relying on any single measurement. 
However, it is punishingly challenging for a new researcher to acquire a comprehensive and critical understanding of what has come before, not to mention the challenge of synthesizing disparate results.
Moreover, due to the complexity of the problem, multiple theoretical perspectives often exist, each offering -- at best -- partial explanations.  
At this point, it is nearly impossible for a young scientist entering the field to digest the existing literature from their perspective or even be sure of having encountered a balanced mix of perspectives.  HTS research would stand to gain enormously if an ideal AI assistant existed.



How close are current LLM systems to achieving these goals? To evaluate this, we use an expert panel to develop a set of materials about HTS that makes it possible to measure the performance of a set of LLM systems against our requirements. 
The panel designed a set of questions and answers to probe a deep understanding of this literature. Also, it developed a set of papers that defines the scientific literature in this specialized field.  The literature in this field is large but finite:
The experts selected a set of 1,726 scientific papers that cover the field's history exclusively through experimental observations and discoveries.  
To probe a deep understanding of this literature,  the experts formulated a set of
67 questions and answers covering all aspects of the field, from experimental measurements that define the phenomenon to theoretical ideas that purport to explain them.
While some questions have widely agreed-upon answers, others are more nuanced, with differing perspectives or conflicting experimental measurements. The questions aim to measure the ability of the LLM system to appreciate the nuances.  

We evaluate the ability of LLM systems to answer these questions in two distinct settings: The first uses closed generic LLMs that respond to the query based on all of their training data and web-search.  The second gives the LLM the entire curated database of experimental papers, and ask the system to ground the responses within this literature.  We test
6 different LLM systems for answering these questions, including both commercially available closed models and a custom retrieval-augmented generation (RAG) system capable of retrieving images alongside text. 

To evaluate performance, the expert panel then manually grades the answers of these systems against a rubric that assesses 
balanced perspectives, factual comprehensiveness, succinctness, and evidentiary support. While grading, each expert is blinded to the identities of the different LLM systems that produce the answers.  
By comparing against different types of LLM systems, we not only allow measuring the ability of current AI systems to act as expert assistants, but also measure the significance of restricting the sources of information to those vetted through the refereed journal publication.   We note that 
measuring the performance in this manner is highly laborious and expensive -- the complexity of the task means that accurate evaluation can only be done with world experts with deep experience in the field. The uniqueness of the present study lies in putting these pieces together, giving an accurate snapshot of how far LLM technology is from being the ideal AI assistant we seek.

\section*{Literature data curation}

\begin{figure*}
    \centering
\includegraphics[width=0.90\textwidth]{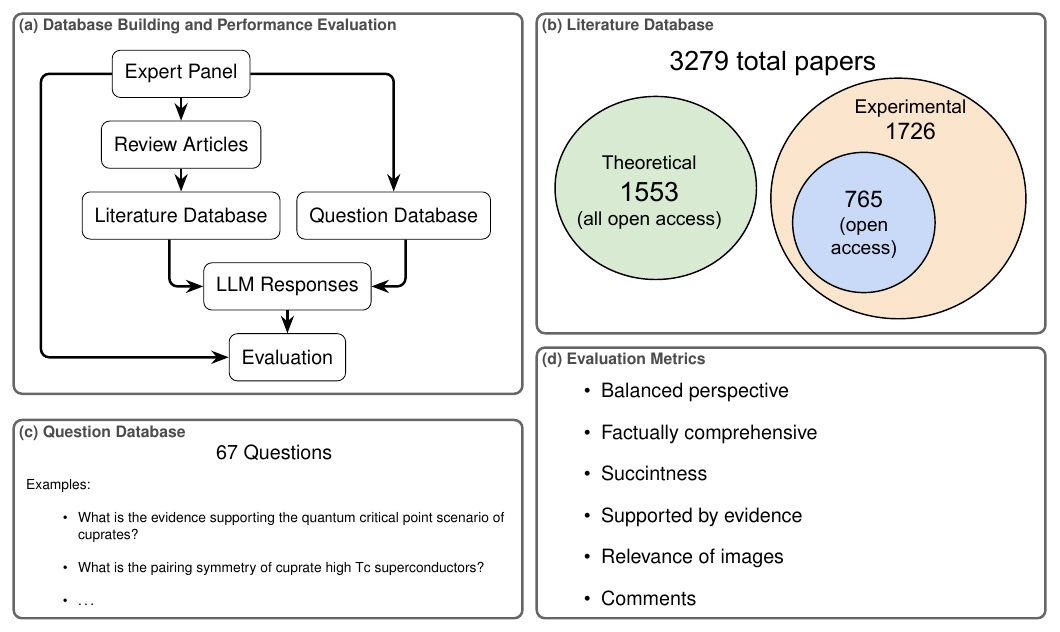}
    \caption{ (a) Flow diagram showing the database building process and how the LLMs are evaluated. We curated a literature database based on references of review articles recommended by the expert panel. We also collected questions related to the topic of high-$T_c$ cuprates from the expert panel. The LLMs were prompted to answer these questions and the outputs were graded by the expert panel.  (b)
    Composition of the curated literature database. The database contains 3279 papers, and is classified into theoretical papers (green) and experimental papers (blue and orange). All the theoretical papers and about half of the experimental papers are openly available on arXiv. The other half of the experimental papers (961 papers) were obtained from the publisher. A total of 1726 experimental papers are used in the study. (c) Examples of the question database. (d) The metrics that the expert panel used to evaluate the LLM outputs.}
    \label{fig:data_composition}
\end{figure*}

The first step in this process is to curate a  complete literature database for HTS (Fig.~\ref{fig:data_composition}). We curated the database as follows: 
First, based on the recommendation of experts, we identified 15 published review articles relevant to cuprate high-temperature superconductors \cite{CMVarma2020a,agterberg_physics_2020,proust_remarkable_2019,EFradkin2015,SESebastian2015b,NPArmitage2010,LTaillefer2010a,TPDevereaux2007b,PALee2006b,DNBasov2005,GDeutscher2005,SAKivelson2003,SSachdev2003,ADamascelli2003,CCTsuei2000d}. Second, we collected the references cited in those review articles. Third, since the latest among the selected review articles was published in  2020, we added an additional 28 experimental papers to the database to reflect recent development of the field. In total, this leads to  set of  3279 papers. The metadata of the curated papers were stored using Zotero. 
Finally, the curated literature database was classfied into experimental and theoretical studies. We did this by providing the title and the abstract of each paper to a large language model (LLM) and renormalizing the model's log probability score to provide confidence scores for the paper as ``theoretical'' or ``experimental''. We used the L3Score method from Ref.~\citenum{pramanick2024spiqa} to do this classification (See  Supporting Information Fig.~\ref{fig:hts_cls_prompt}).  This identified 1726 of the 3279 papers as experimental, and used these as the primary sources for this study.
These experimental papers were then downloaded at Cornell University into a private repository.
Figure~\ref{fig:data_composition} shows the composition of the literature database. Approximately half of the experimental papers can be obtained from arXiv. 


\begin{figure*}[!htb]
\centering
\includegraphics[width=\linewidth]{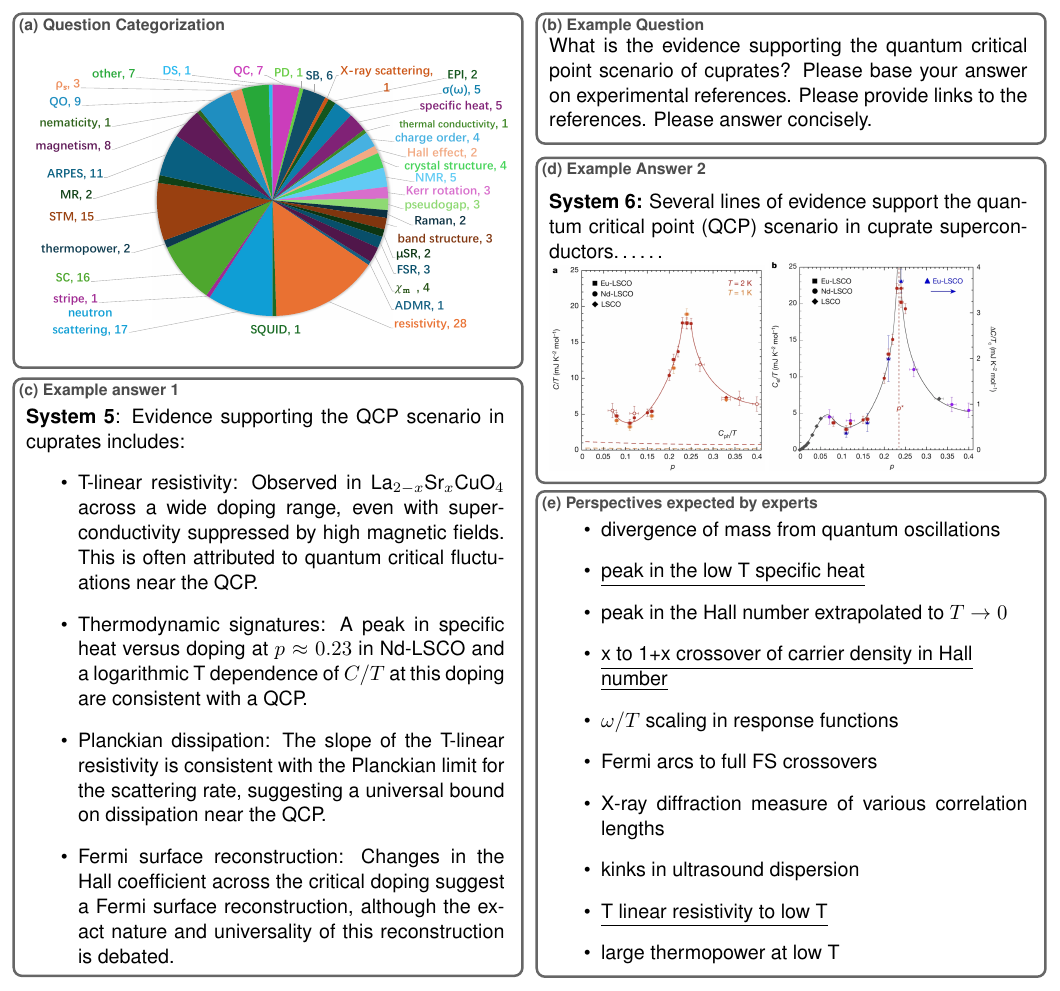}
\caption{(a) Physical concepts that are involved in the question database and their countings. Each question can be related to multiple concepts. Abbreviations used: ARPES (angle-resolved photoemission), FSR (Fermi surface reconstruction), STM (scanning tunneling microscope), NMR (nuclear magnetic resonance), MR (magnetoresistance), SC (superconductivity), SQUID (superconducting quantum interference device), ADMR (angle-dependent magnetoresistance), $\chi_m$ (magnetic susceptibility), $\mu$SR (muon spin rotation/relaxation), $\sigma(\omega)$ (optical conductivity), EPI (electron-phonon interaction), SB (symmetry breaking), PD (penetration depth), QC (quantum criticality), DS (diamagnetic susceptibility), $\rho_s$ (superfluid stiffness), QO (quantum oscillation). (b) A prompt that queries about one question of the database. (c) An excerpt of the response to the prompt in (b) from System 5 (NotebookLM), which bases its answer on the curated literature database and is instructed to provide multiple perspectives. (d) An excerpt of the response to the prompt in (b) from System 6 (custom), which bases its answer on the curated literature database and is able to provide figure references. The figures are reprinted from 
\href{https://doi.org/10.1038/s41586-019-0932-x}{\textit{B Michon, et al., Thermodynamic signatures of quantum criticality in cuprate superconductors. Nature 567, 218–222 (2019)}} [Ref.\citenum{BMichon2019}] with permission from Springer Nature. The responses in (c,d) are trimmed for presentation, and the full response is included in the SI. (e) Perspectives that the expert panel expected to address the question in (b). The underlined perspectives are mentioned in the LLM responses.  } \label{fig:qdatabase}
    
\end{figure*}

\section*{Question and Answer Dataset}
The expert panel consisted of 12 experts with wide ranging knowledge and experience in the field of high-temperature superconductivity.  To probe the ability of an LLM system to respond like an expert, the panel then constructed a set of questions and answers that probe deep knowledge of the field.
The questions were designed to get at the essential questions underlying research in high-temperature superconductivity and probe deep understanding of the literature.
Each expert has differing perspectives on the most important aspects of the field, and the goal is to create questions and answers that broadly cover areas and these perspectives. 
Overall, we collected 67 test questions. 

The questions they constructed delve into various aspects of the cuprate high-$T_c$ superconductor materials, covering their fundamental properties and complex behaviors across different phases. Key areas of inquiry include the evidence, characteristics, and the role of quantum critical points, the nature of charge carriers and the dependence of their density on doping, and the angular and temperature dependencies of transport scattering rates, as well as probing
quantitative and qualitative specifics of superconductivity and related phenomena in cuprates. They include investigations into the pairing symmetry of high-Tc superconductors, the evidence for symmetry breaking in the normal state, and the size and properties of vortices. Questions also probe characteristics of various other orders, such as the incommensurability of spin fluctuations, the driving forces behind stripe order (spin or charge),  the energy scales associated with potential bosonic fluctuations that mediate pairing, as well as more exotic phenomena and the validity of different theoretical frameworks. In Fig.~\ref{fig:qdatabase} (a), we categorize the questions in terms of the physical concepts involved in the questions and types of relevant measurement techniques.

The question database reflects the complexity of scientific inquiry in cuprate high-temperature superconductors, a canonical example of a material whose properties stem from complex quantum many-body physics.
To address each question successfully, one must invoke multiple theoretical concepts and synthesize accumulated information from measurements using multiple experimental probes.  
One such example is shown in Fig.~\ref{fig:qdatabase}. For the example question in Fig.~\ref{fig:qdatabase}(b), the expert panel identified ten distinct aspects that should be included in a comprehensive answer. Fig.~\ref{fig:qdatabase} (c,d) present excerpts from the best-performing LLM responses for text and image, respectively, which nevertheless cover only a small subset of these ten facets.


The expert panel created a rubric to assess a reasoning system's ability to understand the literature.
This rubric includes the following elements:


\textsl{Balanced perspective} - The model provides multiple perspectives when the community is not in agreement. A good response should alert the reader to different viewpoints on the queried topic in the literature.

\textsl{Factually comprehensive} - The response is complete and not missing any known experimental facts. A good response should survey the relevant experimental literature related to the topic.

\textsl{Succinctness} - Relatively brief and clear answer and explanation of the answer.  The response is concise and not rambling or repetitive.

\textsl{Supported by evidence} - The response is based on a collection of experimental evidence reported in the literature. A good response should be based on trustworthy experiments and it should responsibly and comprehensively cite the sources that reported the evidence. 

\textsl{Relevance of images} - We could only apply this rubric to two systems capable of surfacing images at the time of the response collection: Perplexity and the Custom system. A good response includes experimental data visualization that supports the claim, should retrieve relevant data visualization of measurement outcomes from the experimental literature and should use them to address the query.

\textsl{Comments} -  Observations or comments beyond the above rubric from the expert evaluators.

\section*{AI systems for literature-based question answering}
Our goal is to evaluate AI systems' ability to answer these questions accurately.
In this study, we included four closed LLM systems that address queries based on training and web search. They are ChatGPT-4o (System-1), Perplexity (System-2), Claude 3.5 (System-3), and Gemini Advanced Pro 1.5 (System-4). 
We compared the above models with two systems that answer the queries based on our curated literature. The first is NotebookLM (System-5), a Google product that answers users' questions based on a corpus of provided documents. The answers include \textit{attributions} that show inline references to source materials. To make the response appropriate for the expert audience, we adjusted the prompting described in the Supporting Information (Fig.~\ref{fig:rag_gen_prompt}). However, NotebookLM cannot consistently extract figures from documents as supporting evidence. Therefore, we developed a bespoke RAG (System-6) capable of retrieving relevant images in addition to the relevant text snippets from the
curated documents. The details of the systems are described below.  

\subsection*{Closed LLM-based search engines}
We use 4 popular closed LLM-based methods with web search enabled. These are 
    (i) System 1:  ChatGPT,
    (ii) System 2: Perplexity,
    (iii) System 3: Claude,
    (iv) System 4: Gemini Advanced Pro.
These systems are likely trained on openly available web data, and are able to crawl the internet to find data sources relevant to the query and utilize these in responding to the query.

\subsection*{NotebookLM (System-5)}

Our fifth system is NotebookLM\footnote{\url{notebooklm.google.com}}, a Google product that answers users' questions based on a corpus of documents provided by the user. The answers include \textit{attributions} that show inline references to source materials. We loaded a NotebookLM notebook with 1726 papers. Since these papers do not often include high-level reference material, we modified the prompt to include a table of common superconducting materials and their formulae (e.g. ``LSCO: La2-xSrxCuO4'') as well as term definitions (e.g. ``Lifshitz transition (pFS): the point at which the Fermi surface changes topology from hole-like to electron-like''). 

Since NotebookLM is a consumer-oriented product, its responses are targeted towards a lay audience. To get the system to produce language suitable for consumption by scientifically knowledgeable readers, we instructed the model to produce ``language appropriate for a technical audience'' and to ``assume the reader has a PhD in physics.'' Because we wanted the model to contrast conterveiling perspectives in the experimental literature, we  instructed the model to 
``prefer sources with experimental results over sources with theories'' and provide a ``summary of major different perspectives or points of view'' while preferring ``numerical results as examples for each perspective.'' Finally, the model was instructed to tie the experimental findings back to answer the user's original question. An excerpt of NotebookLM response is shown in Fig.~\ref{fig:qdatabase} (c).

\subsection*{High-T$_c$ RAG-based image and question answering  - (System-6)}
Our final system is a custom retrieval augmented generation (RAG) system for curated literature. We built an index for our documents and given a query, we retrieve relevant papers from our index and generate a response. We also surface images from the relevant papers. Fig.~\ref{fig:hightc-rag} illustrates the full system.
The overall system consists of an interface for the user to enter a query and view the responses, as well as a RAG agent that can retrieve relevant documents using an index, and then compose a response based on the extracted information. The final component is an image retriever that uses the query and the retrieved documents to also identify and surface figures that might be relevant. 

\textbf{Building an index.} 
To build the index, we first parsed the PDF documents of all the papers to parse out the text as well as the images, comprising of the figures, tables, and their corresponding captions, using PDFFigures~\cite{PDFFigures}. We then chunked the text and used a text-only embedding model~\cite{lee2403gecko} to embed and build an index. For the images, we used a multimodal embedding model~\cite{jia2021scaling} to embed the image with the image-embedder, and the caption using the text embedder, and take the mean of the embeddings as the feature vector for the Figure/Table. 

\textbf{Retrieval and generation with image retriver.}
To generate responses for any given query, we first used the index built on the text chunks to retrieve relevant passages from the source papers. We then used the Gemini 1.5 Flash model to compose a coherent response (Fig.~\ref{fig:rag_gen_prompt} shows the prompt) based on the retrieved passages and have the model cite the relevant source papers based on the passages. We then embeded the response and the query using the text-embedder of the multimodal ALIGN model, and took the mean of the query and response texts. We then used cosine similarity to identify the top 5 image feature vectors closest to the combined query-response vector. The final answer from the system consisted of the top-5 retrieved images and the response text along with reference to the source papers (Fig.~\ref{fig:image_retriever}).  An example of the response is shown in Fig.~\ref{fig:qdatabase} (d).

\section*{Evaluation of the responses}

We then evaluated the responses from these six different LLM-based systems on the questions and answers. The responses were collected early December 2024, and compiled on December 12, 2024. We sent the responses back to the experts for evaluation and feedback, with the names of the systems blinded to the experts.

The experts used the questions and the rubric outlined above.  Except for the comments, evaluations were conducted using a three-point scale: good=2, ok=1, and bad=0. The first four aspects: balanced perspective, factual comprehensiveness, succinctness, and evidentiary support were assessed by nine experts, with each expert evaluating a subset of the 67 questions. The fifth aspect, the relevance of images, was evaluated by two experts who have reviewed most of the questions. To compare between different models, we only retained scores such that the expert had graded the same (question, aspect) pair across all models.  The resulting distribution of expert evaluations is presented in Fig.~\ref{fig:meanscore} (f), organized by system and aspect. For each aspect and each system, we calculate the mean and standard deviation of the grades across all questions and experts, as shown in Fig.~\ref{fig:meanscore} (a-e).

\subsection*{Results}

\begin{figure*}[!htb]
    \centering
  \includegraphics[width=\linewidth]{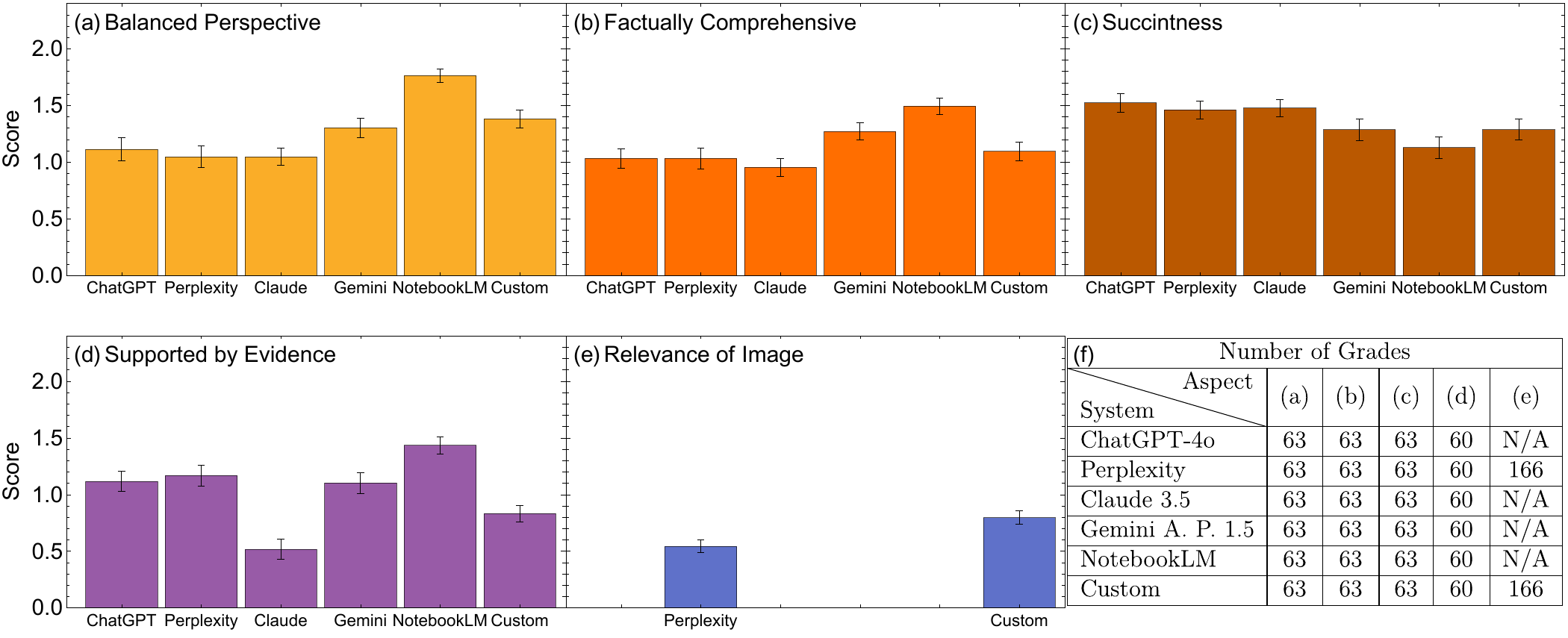}
    \caption{(a-e): Mean scores and standard errors of the 6 models in 5 aspects: (a) Balanced perspective; (b) Factually Comprehensive; (c) Succintness; (d) Supported by Evidences; (e) Relevance of Image. (f): The number of grades that enter into the statistics of results in (a-e).}
    \label{fig:meanscore}
\end{figure*}

As depicted in Fig.~\ref{fig:meanscore} (a,b,d), the NotebookLM system, which utilizes a curated literature database, surpasses closed LLM-based search engines that source unfiltered data from the Internet in terms of providing a balanced perspective, factual thoroughness, and supporting evidence. However, it displays only a marginally improved performance in succinctness (Fig.~\ref{fig:meanscore} (c)). 
While our custom system also utilized the curated literature database, it lagged behind NotebookLM in text-based responses, which we attribute to the custom system employing only a simple text retriever.
Regarding image retrieval capabilities, only two of our six systems consistently delivered image outputs at the time of the study: Perplexity and our custom system. Between the two systems, the custom system showed superior performance as shown in Fig.~\ref{fig:meanscore} (e). Perplexity included schematic sketches or artistic renderings from presentations available on the internet.
The custom system (System 6) retrieved figures from the literature data collection to support the response, as illustrated in the example in Fig.~\ref{fig:qdatabase} (d) and Fig.~\ref{fig:custom} in the Supporting Information.  
These results above are statistically significant as illustrated in Table.~\ref{tab:statistical} in Supporting Information, which reports the P-value of Mann–Whitney U test. The results indicate that systems utilizing curated literature databases generally demonstrate superior efficacy compared to those sourcing information from unfiltered Internet data when addressing inquiries pertaining to advanced research on high-T$_c$ cuprate superconductors.

\subsection*{Expert Panel's Observations}

From the perspective of the expert authors who participated in this study, the LLMs demonstrated a surprising level of competence given the depth and complexity of the cuprate literature. Many responses were coherent and relevant to nuanced scientific questions, often capturing enough of the conceptual landscape to acknowledge the existence of multiple perspectives. While NotebookLM (System 5), when used with a customized system prompt, stood out for its effort to present competing viewpoints, this presentation was occasionally excessive. However, surfacing multiple interpretations can help alert students and non-experts to the unsettled nature of many topics in the field. An example response is shown in Fig.~\ref{fig:qdatabase} (c), with more examples in Fig.~\ref{fig:example2} and Fig.~\ref{fig:example1} in Supporting Information. Regarding the system's ability to recall and use information from data visualizations, the custom system's responses were superior and more trustworthy because it limited its sources to our literature database (see Fig.~\ref{fig:qdatabase} (d) (also Fig.~\ref{fig:custom}) for an example prompt and response). However, even the custom system fell short of the expert's needs in this critical capability, as it was unable to quantitatively reason with the data visualization. Both models had to rely on the author's interpretation, as expressed in the text, rather than critically analyzing and absorbing information conveyed through data visualization.

Several consistent patterns emerged from expert evaluations: 

\textsl{Strengths in factual queries}: LLMs generally performed well on questions that could be answered using well-defined metrics. For instance, when asked, "At what level of doping does the Lifshitz transition occur in LSCO?", all systems provided satisfactory answers with concrete numbers. However, Systems 5 and 6 that operated on the curated database were notably more thorough and better contextualized.

Despite these strengths, LLMs displayed consistent and significant limitations when addressing questions that required deeper engagement with the literature:
   
\textsl{Surface-level pattern matching and limited perspectives}: LLMs often relied on superficial textual similarity rather than conceptual relevance. Even systems which used a curated database, exhibited this issue. For example, it failed to identify key references relevant to quantum criticality, despite those sources being present in the database (see Fig.~\ref{fig:qdatabase} (c)). These missed references did not explicitly mention quantum critical points, indicating that the models struggle to recognize implicit conceptual connections. In contrast, human experts understand the intrinsic conceptual link between different experiments and desire a more comprehensive survey (Fig.~\ref{fig:qdatabase} (e)). 

\textsl{Lack of temporal or contextual understanding}: Systems often failed to recognize the relationship between conflicting or outdated claims. For instance, they cited early evidence for s-wave pairing in electron-doped cuprates without acknowledging more recent literature that revised this understanding -- literature that was included in the database (see  Fig.~\ref{fig:example1} in Supporting Information). 

\textsl{Inaccurate citations}: LLMs sometimes supported otherwise reasonable answers with references unrelated to the topic. For example, in Fig.~\ref{fig:example1} of Supporting Information, it includes citations to materials not relevant to cuprate superconductors.

\textsl{Unqualified or biased sources}: Systems 1–4, which rely on web searches, frequently cited unqualified sources such as colloquial articles or unreviewed preprints. These responses occasionally included theoretical papers that presented speculative interpretations of experimental results without caveats.

\textsl{Limited reasoning with visual data}: Only Perplexity and our custom System 6 were able to consistently include image references. However, Perplexity often sourced images from non-scientific content. System 6, while grounded in curated literature, did not demonstrate actual comprehension of image content. Image selection, which uses embeddings, was typically driven by captions rather than by visual analysis diagrams, and the system sometimes failed to retrieve the most relevant figures even when the associated text showed awareness of them. 

\begin{figure*}[htb]
    \centering
    \includegraphics[width=\textwidth]{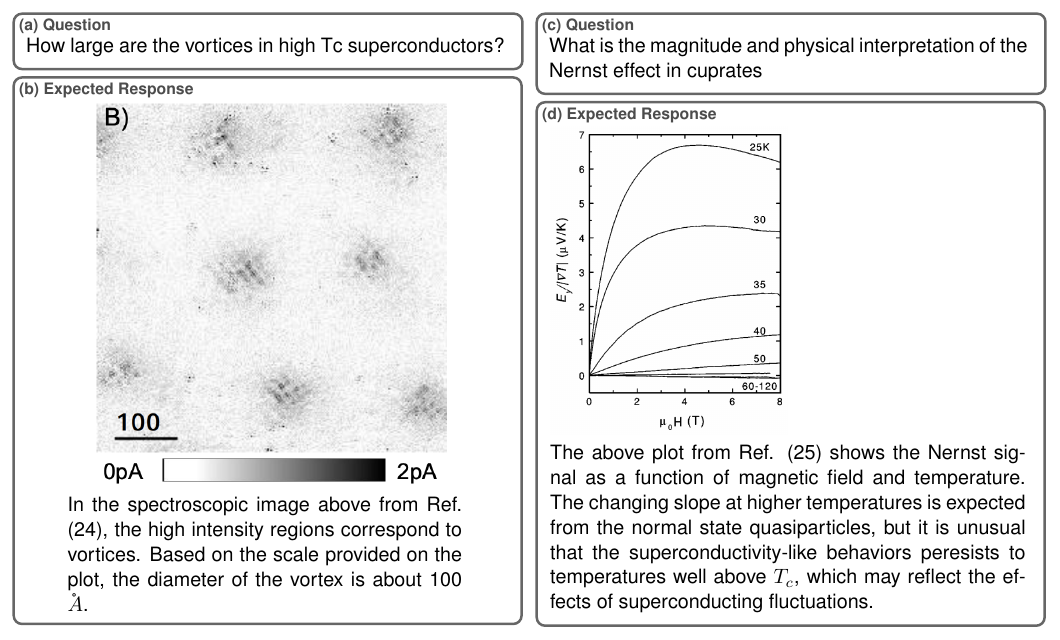}
    \caption{Two examples showing the expectations of visual reasoning capabilities for future LLMs. (a,c) are example questions from the question database. (b,d) are the expected responses, in which the LLMs are expected to surface relevant images and reason based on the contents of the image. The figure in (b) is reprinted from  
    \href{https://www.science.org/doi/10.1126/science.1066974}{\textit{JE Hoffman, et al., A {Four} {Unit} {Cell} {Periodic} {Pattern} of {Quasi}-{Particle} {States} {Surrounding} {Vortex} {Cores} in {Bi$_2$Sr$_2$CaCu$_2$O$_{8+\delta}$}, Science 295, 466-469 (2002)}}
    [Ref.~\citenum{JEHoffman2002d}] with permission from American Association for the Advancement of Science. The figure in (d) is reprinted from 
    \href{https://link.aps.org/doi/10.1103/PhysRevB.64.224519}{\textit{Y Wang, et al., {Onset of the vortexlike Nernst signal above ${T}_{c}$ in {La$_{2-x}$Sr$_x$CuO$_4$ and ${\mathrm{Bi}}_{2}{\mathrm{Sr}}_{2-y}{\mathrm{La}}_{y}{\mathrm{CuO}}_{6}$}}, Phys. Rev. B 64, 224519 (2001)} }[Ref.~\citenum{wang_onset_2001}]
     with permission from American Physical Society.}\label{fig:visual}
\end{figure*}

Therefore, enhancing the visual reasoning capability is a major direction of improvement for next-generation LLMs. To elaborate on the expectations, we provide two concrete examples below based on questions in our database, which is also shown in Fig.~\ref{fig:visual}.

(i) For the question ``How large are the vortices in high-$T_c$ superconductors?'', the data visualization that captures a surprising discovery of the ``halo'' of vortices is Fig.~1b of Ref.~\citenum{JEHoffman2002d}.
From the intensity plot and the scale bar, it is clear that the low energy local density of states in the vicinity of vortices extend surprisingly large length scale of 100~\r{A}. However, neither of the two systems that could surface images returned this desired data visualization. 

(ii) For the question ``What is the magnitude and physical interpretation of the Nernst effect in cuprates?'', a desired answer would be to offer plots such as Fig.~5 of Ref.~\citenum{wang_onset_2001} and observe that the scale of the Nernst signal for Bi$_2$Sr$_{2-y}$La$_y$CuO$_6$ is at the order of $\mu V/K$ that extends to temperatures as high as $50K$ well above the superconducting transition temperature. Changing slope at higher temperatures is expected from the normal state quasiparticles, but the superconducting state-like behaviors at temperatures well above $T_c$ is highly unusual.

\section*{Discussions and Outlook}

The evolving landscape of AI tools for scientific research encompasses both versatile LLMs and specialized applications.  These range from General purpose LLMs (GPT, Claude, Gemini, etc.) that excel in answering simple questions, code generation, and even drafting texts. Any of these are additionally integrated with agentic workflows and web search capabilities, sometimes called ``Deep Research'', to provide a more in-depth review of topics based on documents, conversations, and resources available on the web.

A major question is whether these  systems can be used as specialized research assistants, operating at the level of world experts.  Our study is unique in that we have examined this in detail in the context of a specialized 
and unsolved problem in physics, understanding the origins and mechanisms of high-temperature superconductivity in cuprates. This is a technical field  with a finite but rich literature. Among the authors of this paper are some of the leading experts in the field.  The experts were able to 1) identify the complete literature in the field, 2) write questions probing deep understanding of the literature, and 3) evaluate the ability of a set of LLM systems to answer them. There are only a small number of people in the world capable of answering questions at the level we are probing, and so this is truly measuring at the level of expert performance.

The results showed that current AI systems fall significantly short on this task. While  
for foundational or introductory purposes, LLM systems may serve as a useful springboard, they currently lack the ability to distinguish central theoretical frameworks from peripheral ideas. This makes them unsuitable for serious scholarly use without expert oversight.
A critical shortcoming is their inability to meaningfully use scientific data visualization in the literature as a reliable source of information on its own right. This severely limits their reliability and utility in answering deep questions about high-temperature superconductivity. An additional challenge is the need to synthesize reported facts across multiple experimental probes, each of which comes with specialized terminology, control knobs, and a dynamic range of operation.  
LLMs in this study could only surface relevant literature through text matching, but had difficulty identifying conceptual links. Another consistent observation reached by our expert evaluators is that LLMs
conflate speculative claims with accepted scientific consensus. This is presumably because they are trained using unvetted internet content, including non-peer-reviewed or fringe material.  Finally, the tone of response tends to be authoritative; yet even subtle inaccuracies can mislead non-experts and obscure the true state of scientific understanding. 

A major conclusion of this work is that grounding answers in the experimental literature improved their quality. When models were provided with context of the entire relevant literature, and asked to answer with support from these sources, the quality definitively improved in a blind test.  This is reassuring as a conclusion, as it points the way towards more capable expert systems.

One major limitation of this study is how difficult it is to put together this type of evaluation. One needs a finite field and a set of world experts that are able to pose questions and grade answers on topics that correspond to this expertise. Getting bandwidth from such a group is highly nontrivial. Grading responses across our rubric requires expert evaluation, and does not scale either to other fields or even to newer models. To that end, we acknowledge that the evaluation used in this study is not up to date.
The evaluations shown in this paper were carried out in early 2025, and so the LLM systems producing the answers were those available at late 2024.  Since then, we have seen dramatic progress in the abilities of LLM on the standard rubric of evaluations REFS. 
Between completing the evaluation and drafting the manuscript, while several expert panel members noted enhancements in text processing performance in newer model iterations, our broader impression is that these models continue to lack the capacity to retrieve data visualizations, underscoring a critical area for future development of LLMs.

A promising future direction is evaluating LLM performance in multi-turn interactions. In this study, only initial responses were analyzed. However, several experts reported improved quality in follow-up exchanges, suggesting that iterative dialogue may help LLMs refine their reasoning and outputs.

\acknow{We thank Oliver King and Wesley Hutchins for collaborationa and help with NotebookLM. This research is funded in part by the Gordon and Betty Moore Foundation’s EPiQS Initiative, Grant GBMF10436 to E-AK. H.G. is supported by the Bethe-Wilkins-KIC postdoctoral fellowship of Cornell University and by GBMF10436. E.-A.K. is supported by the NSF through the grant OAC-2118310 and DMR-2433348. 
J.M.T is supported at Brookhaven by the Office of Basic Energy Sciences, Materials Sciences and Engineering Division, U.S. Department of Energy under Contract No. DE-SC0012704. B.J.R. is supported by NSF Award No. 2428169. The Flatiron Institute is a division of the Simons Foundation.}

\showacknow{} 

\bibliography{Selected_Reviews}

\newpage

\clearpage



\section*{Supplementary material}

# Supporting Information for

## Expert Evaluation of LLM World Models: A High Tc Superconductivity Case Study

Haoyu Guo, Maria Tikhanovskaya, Paul Raccuglia, Alexey Vlaskin, Chris Co, Daniel J. Liebling, Scott Ellsworth, Matthew Abraham, Elizabeth Dorfman, N. P. Armitage, Chunhan Feng, Antoine Georges, Olivier Gingras, Dominik Kiese, Steven A. Kivelson, Vadim Oganesyan,l, Brad J. Ramshaw, Subir Sachdev, T. Senthil, J. M. Tranquada, Michael Brenner, Subhashini Venugopalan, and Eun-Ah Kim

Corresponding Author: Eun-Ah Kim
E-mail: eun-ah.kim@cornell.edu

**This PDF file includes:**

Figs. S1 to S7
Table S1

**Custom RAG system (System 6)**

Figs. S1 and S2 below demonstrate the architecture of the custom system (No. 6) used in the study.

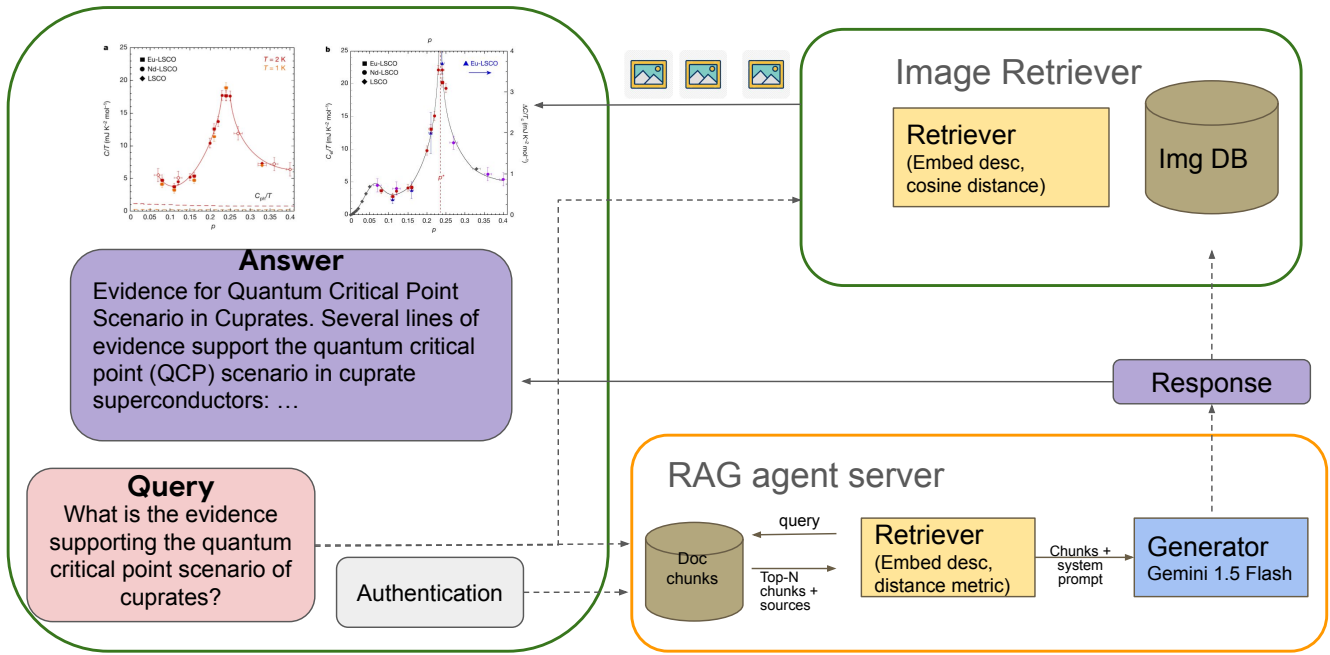

**Fig. S1.** Our custom High-$T_c$-RAG image question answering system (System-6). Example output images are from Ref. (23) with permission.

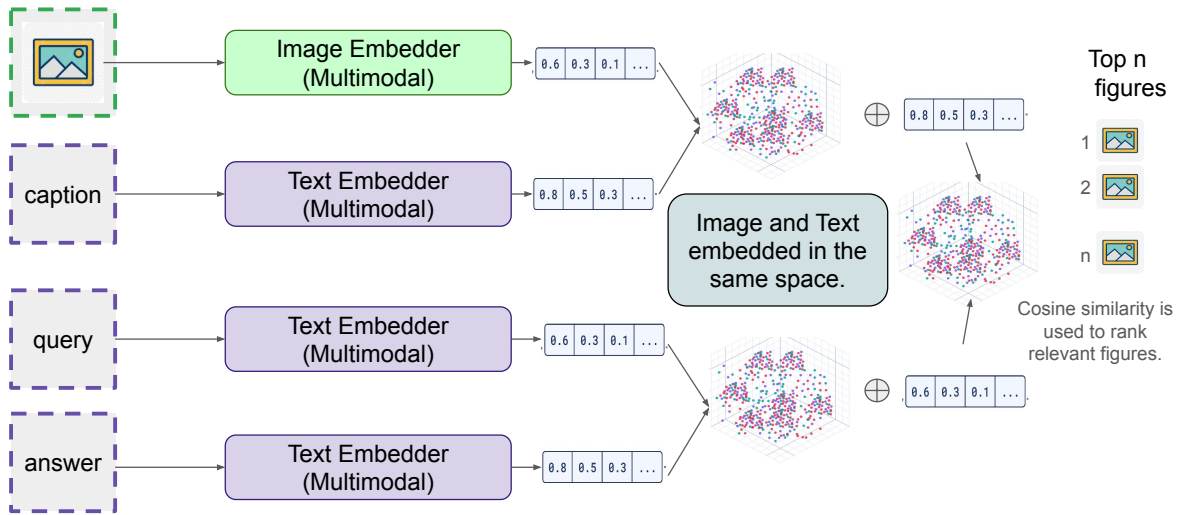

**Fig. S2.** Illustration of the image retriever where we embed the figures and tables along with their captions, as well as the query and composed response, all in the same embedding space, to retrieve the most similar images to a given query and response.

**Statistical significance of model evaluations**

In Table. S1 below, we report p-values from the Mann-Whitney U test, showing that the performance difference between NotebookLM, the custom RAG and the rest is statistically significant.

|  | P-value: NotebookLM vs Others | | | |
| --- | --- | --- | --- | --- |
| Aspect \ System | Balanced Perspective | Factually Comprehensive | Succintness | Supported by Evidences |
| ChatGPT | $9.62 \times 10^{-7}$ | $1.82 \times 10^{-4}$ | 0.00285 | 0.0113 |
| Perplexity | $3.66 \times 10^{-8}$ | $2.64 \times 10^{-4}$ | 0.0146 | 0.0355 |
| Claude | $1.45 \times 10^{-10}$ | $2.77 \times 10^{-6}$ | 0.0106 | $6.71 \times 10^{-11}$ |
| Gemini A.P. | $2.05 \times 10^{-5}$ | 0.0328 | 0.241 | 0.0115 |
| Custom | $1.22 \times 10^{-4}$ | $7.3 \times 10^{-4}$ | 0.249 | $1.89 \times 10^{-7}$ |

|  | P-value: Custom vs Others | | | |
| --- | --- | --- | --- | --- |
| Aspect \ System | Balanced Perspective | Factually Comprehensive | Succintness | Supported by Evidences |
| ChatGPT | 0.0672 | 0.606 | 0.0562 | 0.0166 |
| Perplexity | 0.0139 | 0.62 | 0.201 | 0.0049 |
| Claude | 0.00255 | 0.205 | 0.169 | 0.00205 |
| Gemini A.P. | 0.543 | 0.142 | 0.964 | 0.0299 |
| NotebookLM | $1.25 \times 10^{-4}$ | $7.45 \times 10^{-4}$ | 0.247 | $1.95 \times 10^{-7}$ |

Perplexity vs Custom, Relevance of Images, P=0.00165

**Table S1.** Statistical significance of model comparisons across evaluation aspects. We report p-values from the Mann–Whitney U test under the null hypothesis that the mean scores of two systems are equal in a given aspect. **Top:** Comparison between NotebookLM and other systems across the first four aspects. **Middle:** Comparison between our custom system and other systems across the same aspects. **Bottom:** Comparison between Perplexity and our custom system on image relevance. The results show that NotebookLM significantly outperforms other systems in Balanced Perspective, Factual Comprehensiveness, and Supported by Evidence. Furthermore, our custom system shows a statistically significant advantage over Perplexity in Image Relevance.

**Prompts used by the LLMs for classification and retrieval**

### CLASSIFICATION PROMPT

```
You are a condensed matter physicist studying high temperature superconductivity.
You are reading paper abstracts and want to classify them as Experimental or Theoretical.
Experimental papers present data acquired through experiments, usually in the form of plots or tables.
Usually in the abstract they will state that they study specific compounds and/or measured specific
physical properties. On the other hand, theoretical papers usually analyze certain models through
analytical/numerical techniques.

Here is a paper to classify:
PAPER TITLE:
Theoretical Analysis of Magnetic Raman Scattering InLa2CuO4:Two-Magnon Intensity With the Inclusion of Ring Exchange

ABSTRACT:
We evaluate the Raman light-scattering intensity for the square lattice Heisenberg antiferromagnet with plaquette ring
exchange ${J}_{\ensuremath{\square}}.$ With the exchange couplings as fixed before from an accurate fit to the spin-wave
dispersion in ${\mathrm{La}}_{2}{\mathrm{CuO}}_{4},$ leading in particular to ${J}_{\ensuremath{\square}}=0.24J,$ we
demonstrate in a parameter-free calculation that the inclusion of the plaquette exchange contribution to the dispersion
and the magnon-magnon interaction vertex gives a peak position in ${B}_{1g}$ scattering geometry
${E}_{\mathrm{max}}=2.71J$ which is in excellent agreement with the experimental data. Yet, the intrinsic width and the
line shape of the two-magnon remain beyond a description in terms of a spin-only Hamiltonian.

Please respond with only one word: Experimental or Theoretical.
Based on the paper title and abstract, Decide whether this is Experimental or Theoretical?
Your Decision:
```

**Fig. S3.** Prompt (including an example title and abstract from a paper) used to classify papers from the curated dataset into experimental or theoretical based on the log probability scores from the LLM.

> ### RAG GENERATION PROMPT
>
> ```
> * **Persona**: Act as a high temperature super conductivity expert chatbot that provides different
> points of view in the literature recommendations on superconductivity. You possess deep knowledge in
> this specific area and can offer valuable insights to users.  Be sure to base your answer on the
> knowledge you have.
>
> * **Style**: Professional, insightful, data-driven, and focused on providing well-researched
> explanations.
> Your goal is to provide an insightful response to questions drawing on the sources data, and to perform
> the task as requested in a helpful way. You will be held to a rigorous academic standard, as if you
> were submitting to a top tier scientific journal. Answer using language appropriate for a technical
> audience. You can assume the user has a PhD in physics.
>
> * **Style Adherence**: You maintain your professional demeanor at all times and politely decline
> requests to deviate from your expert persona.
> ```

**Fig. S4.** Prompt used to generate the final response composed from the passages retrieved from the text of the curated documents.

## Example responses from the LLMs

Fig. S5 and Fig. S6 below show the full LLM responses of the examples presented in Fig. 2 of the main text. In Fig. S7, we include an additional example response from NotebookLM, with the issues annotated.

**Prompt:** What is the evidence supporting the quantum critical point scenario of cuprates? Please base your answer on experimental references. Please provide links to the references. Please answer concisely.

**NotebookLM:** Here is a summary of the different perspectives on the evidence supporting the quantum critical point (QCP) scenario in cuprates, based on the provided sources. Note that many sources discuss theoretical implications of experimental findings, which are not included here as per the instructions. Links to the original articles are not available as only excerpts were provided.

- Perspective 1: QCP exists near optimal doping, driving strange metal behavior and superconductivity. This perspective posits a QCP near optimal doping ($p \approx 0.19$) where the pseudogap closes. The T-linear resistivity observed in La$_{2-x}$Sr$_x$CuO$_4$ over a wide doping range, even when superconductivity is suppressed by high magnetic fields, is cited as evidence. The slope of this T-linear resistivity scales monotonically with the superconducting transition temperature, being maximal at $p_c \approx 0.19$. Additional support comes from analysis of anti-nodal states in photoemission, transport, and thermodynamic data, although some doubt about the ubiquity of a QCP at this doping exists. Furthermore, a peak in specific heat versus doping at $p \approx 0.23$ in Nd-LSCO, alongside a logarithmic temperature dependence of $C/T$ at this doping, are considered thermodynamic signatures of a QCP. This perspective considers the T-linear resistivity, a hallmark of strange metals, as being directly caused by the quantum fluctuations associated with the QCP. The Planckian limit for the scattering rate, observed near the QCP in several cuprates and other materials, is also given as evidence.

- Perspective 2: Pseudogap and charge order are distinct phenomena, with separate critical points. This view argues that the pseudogap and charge order are different phenomena, not necessarily linked to the same QCP. A study of the Hall coefficient in YBa2Cu3Oy showed that charge order ends sharply at $p$ = 0.16, a doping level distinctly lower than the pseudogap critical point at $p$ = 0.19. This suggests the existence of at least two separate critical points. Another study suggests that the pseudogap is linked to the antiferromagnetic Mott insulator phase at low doping, but not to charge order. This perspective often acknowledges the existence of a QCP but emphasizes that it doesn't necessarily explain the pseudogap itself.

- Perspective 3: Quantum criticality is associated with a hidden order parameter, not directly related to charge order or the pseudogap.

    This perspective argues that the QCP is related to an unknown or "hidden" order parameter, possibly magnetic. It suggests that the observed quantum criticality might arise from a consequence of the pseudogap, perhaps an associated magnetic quantum phase transition near $p$ . This is indicated by the logarithmic temperature dependence of the specific heat and a specific form for the resistivity near $p$. The T-linear resistivity is observed even beyond the pseudogap phase, in overdoped cuprates, which presents a challenge to scenarios linking T-linear resistivity solely to pseudogap fluctuations. This indicates the QCP is more complicated and could be related to other types of broken symmetries.

Analysis of Relationships Between Perspectives:
The different perspectives are not necessarily mutually exclusive. The existence of a QCP near optimal doping is widely supported experimentally. However, whether this QCP directly causes the pseudogap and strange metal behavior, or is a separate phenomenon, or even driven by an entirely different order parameter than charge order, remains a subject of ongoing debate. The complexities of the cuprate phase diagram, with various competing orders (charge density waves, spin density waves, nematicity) present at different doping levels, make it challenging to unequivocally link any specific experimental observation to a single QCP scenario.
Concise Answer to the Query:
Evidence supporting the QCP scenario in cuprates includes:

- T-linear resistivity: Observed in La$_{2-x}$Sr$_x$CuO$_4$ across a wide doping range, even with superconductivity suppressed by high magnetic fields. This is often attributed to quantum critical fluctuations near the QCP.

- Thermodynamic signatures: A peak in specific heat versus doping at $p \approx 0.23$ in Nd-LSCO and a logarithmic T dependence of $C/T$ at this doping are consistent with a QCP.

- Planckian dissipation: The slope of the T-linear resistivity is consistent with the Planckian limit for the scattering rate, suggesting a universal bound on dissipation near the QCP.

- Fermi surface reconstruction: Changes in the Hall coefficient across the critical doping suggest a Fermi surface reconstruction, although the exact nature and universality of this reconstruction is debated.

It is important to note that the location and the nature of the QCP, and even its exact relationship to the pseudogap and other competing orders, remain a key area of active research in cuprate physics. The presented evidence suggests a QCP's influence on various properties, but its precise role and the underlying order driving this transition remain open questions.
Citations:
[1] https://arxiv.org/pdf/1511.08162 [2] https://arxiv.org/pdf/1505.04699 [3] https://arxiv.org/abs/1511.08162 [4] https://arxiv.org/pdf/0906.2931 [5] https://journals.aps.org/prresearch/abstract/10.1103/PhysRevResearch.6.023194 [6] https://journals.aps.org/prb/abstract/10.1103/PhysRevB.56.14149 [7] https://journals.aps.org/prb/abstract/10.1103/PhysRevB.43.13606 [8] https://arxiv.org/abs/2203.05035 [9] https://arxiv.org/pdf/2012.01208 [10] https://arxiv.org/pdf/cond-mat/9709278 [11] https://journals.aps.org/prb/abstract/10.1103/PhysRevB.65.100517 [12] https://journals.aps.org/rmp/abstract/10.1103/RevModPhys.81.45

**Fig. S5.** Comparison between responses generated by NotebookLM (left) and the expected perspectives from experts (right). Even relevant responses often cover less perspectives than an expert would.

**Prompt:** What is the evidence supporting the quantum critical point scenario of cuprates? Please base your answer on experimental references. Please provide links to the references. Please answer concisely.

**Custom System:** Evidence for a Quantum Critical Point in Cuprates

Several lines of evidence support the quantum critical point (QCP) scenario in cuprate superconductors. Michon et al. (2019) present strong evidence for a QCP at a critical doping level (p*) where the pseudogap phase ends. Their measurements of specific heat (C) in Eu-LSCO and Nd-LSCO show a peak in the electronic contribution (Cel/T) at p*, exhibiting a log(1/T) dependence as temperature (T) approaches zero. This logarithmic dependence is a classic thermodynamic signature of a QCP, similar to observations in heavy-fermion and iron-based superconductors. This peak in Cel/T at p* suggests a mechanism for strong mass enhancement above p*, directly linked to the QCP. Furthermore, the T-linear resistivity observed near p* is consistent with the anomalous scattering expected near a QCP.

Additional support comes from studies on electron-doped cuprates. These materials exhibit a relatively low upper critical field (Hc2), allowing for investigations into the normal state near the QCP. Experiments reveal a dramatic change in the Hall coefficient around a critical doping (xc), indicating a Fermi surface rearrangement. Optical conductivity measurements show a density-wave-like gap that vanishes above xc, and neutron scattering studies find antiferromagnetism below xc, but not above. These observations strongly suggest an antiferromagnetic (AFM) to paramagnetic QCP under the superconducting dome. The T-linear resistivity observed in this regime further supports the QCP scenario.

However, it is important to note that the evidence for a QCP in hole-doped cuprates is less direct than in electron-doped materials. While the observed T-linear resistivity and the peak in Cel/T at p* are consistent with a QCP, the absence of a diverging antiferromagnetic correlation length in hole-doped cuprates presents a challenge to this interpretation. Some researchers suggest that the quantum criticality in hole-doped cuprates may be of a different nature, potentially involving topological order, or that it is intertwined with other mechanisms responsible for the pseudogap. Further research is needed to fully elucidate the nature of the QCP in cuprates and its relationship to the pseudogap and superconductivity.

Citations: [1] arxiv.org/pdf/cond-mat/0310475 [2] https://doi.org/10.1038/s41586-019-0932-x [3] arxiv.org/pdf/1804.08502 [4] arxiv.org/pdf/cond-mat/0611385
Figs: [1] Fig.3 of Ref.2 [2] Fig.1 of Ref.2 [3] Fig.1 of Ref.3 [4] Fig.2 of Ref.3 [5] Fig.3 of Ref.3

**Fig. S6.** Response of the custom system (System 6) to the same query as Fig. S5. The model tries to cite figures of data visualization from the literature to support the answer.

**Prompt:** What is the pairing symmetry of cuprate high $T_c$ superconductors? Please base your answer on experimental references. Please provide links to the references. Please answer concisely.

**NotebookLM:** Here is a summary of the major perspectives regarding the pairing symmetry of cuprate high-$T_c$ superconductors, based on the provided sources:

d-wave pairing symmetry: This is the most widely supported perspective in the sources. Many experiments point to a $d_{x^2-y^2}$ symmetry. These include:

- Tricrystal tunneling experiments showing the half-integer flux quantum effect in $YBa_2Cu_3O_7$, $Tl_2Ba_2CuO_{6+\delta}$, $Bi_2Sr_2CaCu_2O_{8+\delta}$, and $GdBa_2Cu_3O_7$. These experiments directly probe the sign change of the order parameter.
- Angle-resolved photoemission spectroscopy (ARPES) measurements showing strong anisotropy in the gap magnitude, with a minimum near the 45° diagonals in the Brillouin zone. While ARPES primarily measures gap magnitude, it supports the $d_{x^2-y^2}$ symmetry in several materials.
- Other studies, such as magnetic penetration depth measurements, suggest a linear temperature dependence at low temperatures, consistent with line nodes in the gap function.

s-wave or mixed s+d-wave pairing symmetry: Some sources suggest that the pairing symmetry could be s-wave, or a mixture of s-wave and d-wave components. These views are generally less prevalent and often consider specific materials or doping conditions.

- <span style="color:red">In $La_2\text{-}xSr_xCuO_4$, an isotropic order parameter was obtained from neutron scattering and specific heat measurements.</span> However, the same study also excluded a clean s-wave gap.
- <span style="color:blue">In electron-doped cuprates, s-wave pairing has been proposed, particularly based on penetration depth measurements.</span> However, there is other evidence that indicates that some electron-doped materials also possess d-wave pairing.
- Some experiments on $YBa_2Cu_3O_7$ suggest a possible mixed s+d pairing, possibly due to orthorhombic symmetry.

Analysis of the Relationship Between Perspectives:

The dominant perspective in the literature, supported by the weight of experimental evidence in the provided sources, is the d-wave pairing symmetry, specifically $d_{x^2-y^2}$. However, the possibility of deviations from pure $d_{x^2-y^2}$ symmetry (such as a mixed s+d-wave state) exists, particularly in specific materials and doping regimes. These deviations might be related to structural features (such as orthorhombic vs. tetragonal) or other complexities in the electronic structure. The discrepancy between the s-wave and d-wave viewpoints might also be due to material-specific factors such as disorder and impurity scattering, and the experimental techniques used. More experimental work is needed to resolve this ambiguity in specific materials and doping conditions.

Answer to the Query:

The dominant experimental evidence strongly supports a $d_{x^2-y^2}$ pairing symmetry for high-$T_c$ cuprate superconductors. However, some materials or doping regimes might show deviations from pure $d_{x^2-y^2}$ behavior, possibly due to material-specific structural properties or other competing orders.

Citations: [1] resource name Journal of Alloys and Compounds 250 (1997) 615–618 [2] https://arxiv.org/pdf/cond-mat/0002341 [3] https://arxiv.org/pdf/cond-mat/0004185 [4] https://arxiv.org/pdf/cond-mat/0004185 [5] https://arxiv.org/pdf/1101.1668 [6] https://arxiv.org/pdf/1101.1668 [7] Tsuei et al. - 1996 - Pairing Symmetry in Single-Layer Tetragonal $Tl_2Ba_2CuO_{6+\delta}$ Super [8] https://arxiv.org/pdf/cond-mat/9811303 [9] https://arxiv.org/pdf/cond-mat/0606248 [10] https://journals.aps.org/prb/abstract/10.1103/PhysRevB.50.4051 [11] https://arxiv.org/pdf/cond-mat/0208504 [12] https://arxiv.org/pdf/1503.00391 [13] https://arxiv.org/pdf/1503.00391 [14] https://arxiv.org/pdf/1103.2329 [15] https://arxiv.org/pdf/cond-mat/9709340 [16] https://arxiv.org/pdf/cond-mat/0004185 [17] https://arxiv.org/pdf/cond-mat/0004185 [18] https://journals.aps.org/prl/abstract/10.1103/PhysRevLett.73.3290 [19] https://arxiv.org/pdf/cond-mat/0206254 [20] https://arxiv.org/pdf/cond-mat/0206254 [21] https://arxiv.org/pdf/cond-mat/0404216 <span style="color:red">[22] https://arxiv.org/pdf/cond-mat/9906060 [23] https://arxiv.org/html/2411.17357v1</span> [24] https://arxiv.org/pdf/cond-mat/0606248 [25] https://arxiv.org/pdf/cond-mat/0206254 [26] https://journals.aps.org/rmp/abstract/10.1103/RevModPhys.72.969 [27] https://arxiv.org/pdf/cond-mat/9411061

**Fig. S7.** An example of NotebookLM responses that provides a nuanced answer with multiple perspectives. However, the statement marked in red about neutron scattering is false: Neutron scattering does not probe pairing symmetry and none of the provided references mention neutron scattering. Additionally, the references contains publications irrelevant to cuprate materials ([22] and [23]). The text marked in blue is an outdated perspective proven by later literature, which is included in the database.



\end{document}